\documentclass[MSNbibl,nameyear,noautosecdot,dvips]{arxstspdf}
\usepackage{flushend}
\usepackage{stfloats}

\volume{29}
\issue{2}
\pubyear{2014}
\firstpage{227}
\lastpage{239}
\doi{10.1214/13-STS457} 

\makeatletter

\newproclaim{Note}{Note}

\makeatother

\begin{document}
\begin{frontmatter}

\title{On the Birnbaum Argument for the Strong Likelihood
Principle\thanksref{T1}}%
\relateddois{T1}{Discussed in \relateddoi{d}{10.1214/14-STS470},
\relateddoi{d}{10.1214/14-STS471},
\relateddoi{d}{10.1214/14-STS472},
\relateddoi{d}{10.1214/14-STS473},
\relateddoi{d}{10.1214/14-STS474} and
\relateddoi{d}{10.1214/14-STS475};
rejoinder at \relateddoi{r}{10.1214/14-STS482}.}
\runtitle{On the Birnbaum Argument for the Strong Likelihood
Principle}

\begin{aug}
\author[a]{\fnms{Deborah G.}~\snm{Mayo}\corref{}\ead[label=e1]{mayod@vt.edu}}
\runauthor{D. G. Mayo}

\affiliation{Virginia Tech}

\address[a]{Deborah G. Mayo is Professor of Philosophy,
Department of Philosophy, Virginia Tech,
235 Major Williams Hall, Blacksburg, Virginia 24061, USA \printead{e1}.}
\end{aug}

\begin{abstract}
An essential component of inference based on familiar frequentist notions,
such as $p$-values, significance and confidence levels, is the relevant
sampling distribution. This feature results in violations of a principle
known as the strong likelihood principle (SLP), the focus of this paper. In
particular, if outcomes $\mathbf{x}^{\ast }$ and $\mathbf{y}^{\ast }$ from
experiments $E_{1}$ and $E_{2}$ (both with unknown parameter $\theta $)
have different probability models $f_{1}(\cdot),f_{2}(\cdot)$, then even though $%
f_{1}(\mathbf{x}^{\ast };\theta )=cf_{2}(\mathbf{y}^{\ast };\theta )$ for
all $\theta $, outcomes $\mathbf{x}^{\ast }$ and $\mathbf{y}^{\ast }$ may
have different implications for an inference about $\theta $. Although such
violations stem from considering outcomes other than the one observed, we
argue this does not require us to consider experiments other than the one
performed to produce the data. David Cox [\textit{Ann. Math. Statist.} \textbf{29} (1958) 357--372] proposes the Weak
Conditionality Principle (WCP) to justify restricting the space of relevant
repetitions. The WCP says that once it is known which $E_{i}$ produced the
measurement, the assessment should be in terms of the properties of $E_{i}$.
The surprising upshot of Allan Birnbaum's [\textit{J. Amer. Statist. Assoc.} \textbf{57} (1962) 269--306] argument is that the SLP
appears to follow from applying the WCP in the case of mixtures, and so
uncontroversial a principle as sufficiency (SP). But this would preclude the
use of sampling distributions. The goal of this article is to provide a new
clarification and critique of Birnbaum's argument. Although his argument
purports that [(WCP and SP) entails SLP], we show how data may violate the
SLP while holding both the WCP and SP. Such cases also refute [WCP entails
SLP].
\end{abstract}

\end{frontmatter}

\setcounter{footnote}{1}
\section{Introduction}\label{sec1}

It is easy to see why Birnbaum's argument for the strong likelihood
principle (SLP) has long been held as a significant, if controversial,
result for the foundations of statistics. Not only do all of the familiar
frequentist error-probability notions, $p$-values, significance levels and so
on violate the SLP, but the Birnbaum argument purports to show that the SLP
follows from principles that frequentist sampling theorists accept:
\begin{quote}
The likelihood principle is incompatible with the main body of modern
statistical theory and practice, notably the Neyman--Pearson theory of
hypothesis testing and of confidence intervals, and incompatible in general
even with such well-known concepts as standard error of an estimate and
significance level. [\citet{Bir68}, page 300.]
\end{quote}

The incompatibility, in a nutshell, is that on the SLP, once the data $%
\mathbf{x}$ are given, outcomes other than $\mathbf{x}$ are irrelevant to
the evidential import of $\mathbf{x}$. ``[I]t is
clear that reporting significance levels violates the LP [SLP], since
significance levels involve averaging over sample points other than just the
observed $\mathbf{x}$.'' [Berger and Wolpert (\citeyear{BerWol84}), page 105.]%

\subsection{The SLP and a Frequentist Principle of Evidence (FEV)}\label{sec1.1}

Birnbaum, while responsible for this famous argument, rejected the SLP
because ``the likelihood concept cannot be construed so as
to allow useful appraisal, and thereby possible control, of probabilities of
erroneous interpretations'' [\citet{Bir69}, page 128]. That is, he
thought the SLP at odds with a fundamental frequentist principle of
evidence.

\textit{Frequentist Principle of Evidence (general)}: Drawing inferences
from data requires considering the relevant error probabilities associated
with the underlying data generating process.

David Cox intended the central principle invoked in Birnbaum's argument, the
Weak Conditionality Principle (WCP), as one important way to justify
restricting the space of repetitions that are relevant for informative
inference. Implicit in this goal is that the role of the sampling
distribution for informative inference is not merely to ensure low error
rates in repeated applications of a method, but to avoid misleading
inferences in the case at hand [\citet{May96}; Mayo and Spanos
(\citeyear{MaySpa06}, \citeyear{MaySpa11}); \citeauthor{MayCox06} (\citeyear{MayCox06})].

To refer to the most familiar example, the WCP says that if a parameter of
interest $\theta$ could be measured by two instruments, one more precise
then the other, and a randomizer that is utterly irrelevant to $\theta$ is
used to decide which instrument to use, then, once it is known which
experiment was run and its outcome given, the inference should be assessed
using the behavior of the instrument actually used. The convex combination
of the two instruments, linked via the randomizer, defines a mixture
experiment, $E_{\mathrm{mix}}$. According to the WCP, one should condition on the
known experiment, even if an unconditional assessment improves the long-run
performance [\citet{CoxHin74}, pages 96--97].\vadjust{\goodbreak}

While conditioning on the instrument actually used seems obviously correct,
nothing precludes the\linebreak[4]  Neyman--Pearson theory from choosing the procedure
``which is best on the average over both
experiments'' in $E_{\mathrm{mix}}$ [\citet{LehRom05}, page 394].
They ask the following: ``for a given test or confidence procedure, should
probabilities such as level, power, and confidence coefficient be calculated
conditionally, given the experiment that has been selected, or
unconditionally?'' They suggest that ``[t]he answer cannot be found within the model but depends on the
context'' (ibid). The WCP gives a rationale for using the
conditional appraisal in the context of informative parametric inference.%

\subsection{What Must Logically Be Shown}\label{sec1.2}

However, the upshot of the SLP is to claim that the sampling theorist must
go all the way, as it were, given a parametric model. If she restricts
attention to the experiment producing the data in the mixture experiment,
then she is led to consider just the data and not the sample space, once the
data are in hand. While the argument has been stated in various forms, the
surprising upshot of all versions is that the SLP appears to follow from
applying the WCP in the case of mixture experiments, and so uncontroversial
a notion as sufficiency (SP). ``Within the context of what
can be called classical frequency-based statistical inference, \citet{Bir62}
argued that the conditionality and sufficiency principles imply the
[strong] likelihood principle'' [Evans, Fraser and Monette
(\citeyear{EvaFraMon86N3}), page 182].

Since the challenge is for a sampling theorist who holds the WCP,
it is obligatory to consider whether and how such a sampling
theorist can meet it. While the WCP is not itself a theorem in a formal
system, Birnbaum's argument purports that the following is a theorem:
\begin{eqnarray*}
\mbox{[(WCP and SP) entails SLP]}.%
\end{eqnarray*}%
If true, any data instantiating both WCP and SP could not also violate the
SLP, on pain of logical contradiction. We will show how data may violate the
SLP while still adhering to both the WCP and SP. Such cases also refute [WCP
entails SLP], making our argument applicable to attempts to weaken or remove
the SP. Violating SLP may be written as not-SLP.

We follow the formulations of the Birnbaum argument given in Berger and
Wolpert (\citeyear{BerWol84}), \citet{Bir62}, Casella and  Berger (\citeyear{GarJolJon02}) and  \citet{Cox77}.
The current analysis clarifies and fills in important gaps of an earlier
discussion in Mayo (\citeyear{autokey26}), \citet{MayCox11}, and lets us cut through a
fascinating and complex literature. The puzzle is solved by adequately
stating the WCP and keeping the meaning of terms consistent, as they must
be in an argument built on a series of identities.

\subsection{Does It Matter?}\label{sec1.3}

On the face of it, current day uses of sampling theory statistics do not
seem in need of going back 50 years to tackle a foundational argument. This
may be so, but only if it is correct to assume that the Birnbaum argument is
flawed somewhere. Sampling theorists who feel unconvinced by some of the
machinations of the argument must admit some discomfort at the lack of
resolution of the paradox. If one cannot show the relevance of error
probabilities and sampling distributions to inferences once the data are in
hand, then the uses of frequentist sampling theory, and resampling methods,
for inference purposes rest on shaky foundations.

The SLP is deemed of sufficient importance to be included in textbooks on
statistics, along with a version of Birnbaum's argument that we will
consider:
\begin{quote}
It is not uncommon to see statistics texts argue that in frequentist theory
one is faced with the following dilemma: either to deny the appropriateness
of conditioning on the precision of the tool chosen by the toss of a coin,
or else to embrace the strong likelihood principle, which entails that
frequentist sampling distributions are irrelevant to inference once the data
are obtained. This is a false dilemma. \ldots The `dilemma' argument is
therefore an illusion. [Cox and Mayo (\citeyear{autokey17}), page 298.]
\end{quote}

If we are correct, this refutes a position that is generally presented as
settled in current texts. But the illusion is not so easy to dispel, thus
this paper.

Perhaps, too, our discussion will illuminate a point of agreement between
sampling theorists and contemporary nonsubjective Bayesians who concede they
``have to live with some violations of the likelihood and
stopping rule principles'' [Ghosh, Delampady and Sumanta
(\citeyear{GhoDelSam06}), page 148], since their prior probability distributions are influenced by
the sampling distribution. ``This, of course, does not
happen with subjective Bayesianism. \ldots  the objective Bayesian responds that
objectivity can only be defined relative to a frame of reference, and this
frame needs to include the goal of the analysis.'' [\citet{Ber06}, page 394.] By contrast, Savage stressed:
\begin{quote}
According to Bayes's theorem, $P(\mathbf{x}\mathit{\mid }\theta )$ \ldots \linebreak[4]
constitutes the entire evidence of the experiment \ldots [I]f $\mathbf{%
y}$ is the datum of some other experiment, and if it happens that $P(\mathbf{%
x}\mathit{\mid }\theta )$ and $P(\mathbf{y}\mathit{\mid }\theta )$ are
proportional functions of $\theta $ (that is, constant multiples of each
other), then each of the two data $\mathbf{x}$ and $\mathbf{y}$ have exactly
the same thing to say about the value of $\theta $. [\citet{autokey33}, page 17, using
$\theta $ for his $\lambda $ and $P$ for $Pr$.]
\end{quote}

\section{Notation and Sketch of Birnbaum's argument}\label{sec2}

\subsection{Points of Notation and Interpretation}\label{sec2.1}

Birnbaum focuses on informative inference about a parameter $\theta$ in a
given model $M$, and we retain that context. The argument calls for a
general term to abbreviate: the inference implication from experiment $E$
and result $\mathbf{z}$, where $E$ is an experiment involving the
observation of $\mathbf{Z}$ with a given distribution $f(\mathbf{z};\theta)$
and a model $M$. We use the following:
\begin{quote}
$\mathsf{Infr}_{E}[\mathbf{z}]$: the parametric
statistical inference from a given or known $(E,\mathbf{z})$.%
\end{quote}

(We prefer ``given'' to ``known'' to avoid reference to psychology.) We assume
relevant features of model $M$ are embedded in the full statement of
experiment $E$. An inference method indicates how to compute the informative
parametric inference from $(E,\mathbf{z})$. Let
\begin{quote}
$(E,\mathbf{z})\Rightarrow\mathsf{Infr}_{E}[\mathbf{z}]$:  an informative
parametric inference about $\theta$ from given
 $(E,\mathbf{z})$ is to be computed by means of $\mathsf{Infr}_{E}[\mathbf{z%
}]$.%
\end{quote}
 The principles of interest turn on cases where $(E,\mathbf{z%
})$ is given, and we reserve ``$\Rightarrow $%
'' for such cases. The abbreviation $\mathsf{Infr}_{E}[%
\mathbf{z}]$, first developed in Cox and Mayo (\citeyear{autokey17}), could allude to any
parametric inference account; we use it here to allow ready identification
of the particular experiment $E$ and its associated sampling distribution,
whatever it happens to be. $\mathsf{Infr}_{E_{\mathrm{mix}}}(\mathbf{z})$ is always
understood as using the convex combination over the elements of the mixture.

Assertions about how inference ``is to be computed given $(E,%
\mathbf{z})$'' are intended to reflect the principles of
evidence that arise in Birnbaum's argument, whether mathematical or based on
intuitive, philosophical considerations about evidence. This is important
because Birnbaum emphasizes that the WCP is ``not necessary
on mathematical grounds alone, but it seems to be supported compellingly by
considerations \ldots concerning the nature of evidential
meaning'' of data when drawing parametric statistical
inferences [\citet{Bir62}, page 280]. In using ``$=$%
'' we follow the common notation even though WCP is
actually telling us when $\mathbf{z}_{1}$ and $\mathbf{z}_{2}$ \textit{should%
} be deemed inferentially equivalent for the associated inference.

By noncontradiction, for any $(E,\mathbf{z})$, $\mathsf{Infr}_{E}[\mathbf{z}%
]=\mathsf{Infr}_{E}[\mathbf{z}]$.
So to apply a given inference implication means its inference directive is
used and not some competing directive at the same time. Two outcomes $%
\mathbf{z}_{1}$ and $\mathbf{z}_{2}$ will be said to have the same inference
implications in $E$, and so are inferentially equivalent within $E$,
whenever $\mathsf{Infr}_{E}[\mathbf{z}_{1}]=\mathsf{Infr}_{E}[\mathbf{z}%
_{2}]$.

\subsection{The Strong Likelihood Principle: SLP}\label{sec2.2}

The principle under dispute, the SLP, asserts the inferential equivalence of
outcomes from distinct experiments $E_{1}$ and $E_{2}$. It is a universal
if-then claim:
\begin{quote}
SLP: For any two experiments $E_{1}$ and $E_{2}$ with different
probability models $f_{1}(\cdot)$, $f_{2}(\cdot)$ but with the same unknown
parameter $\theta$, if outcomes $\mathbf{x}^{\ast}$ and $\mathbf{y}^{\ast}$
(from $E_{1}$ and $E_{2}$, resp.) give rise to proportional likelihood
functions ($f_{1}(\mathbf{x}^{\ast};\boldsymbol{\theta})=cf_{2}(\mathbf{%
y}^{\ast};\boldsymbol{\theta})$ for all $\theta$, for $c$ a positive constant),
then $\mathbf{x}^{\ast}$ and $\mathbf{y}^{\ast}$ should be inferentially
equivalent for any inference concerning parameter $\theta$.
\end{quote}

A shorthand for the entire antecedent is that $(E_{1},\mathbf{x}^{\ast})$ is
an SLP pair with $(E_{2},\mathbf{y}^{\ast})$, or just $\mathbf{x}^{\ast}$ and
$\mathbf{y}^{\ast}$ form an SLP pair (from $\{E_{1},E_{2}\}$). Assuming all
the SLP stipulations, for example, that $\boldsymbol{\theta}$ is a shared parameter
(about which inferences are to be concerned), we have the following:
\begin{quote}
SLP:  If $(E_{1},\mathbf{x}^{\ast})$ and $(E_{2},\mathbf{y}%
^{\ast}) $ form an SLP pair, then $\mathsf{Infr}_{E_{1}}[\mathbf{x}^{\ast}]=%
\mathsf{Infr}_{E_{2}}[\mathbf{y}^{\ast}]$.
\end{quote}

Experimental pairs $E_{1}$ and $E_{2}$ involve observing random variables $%
\mathbf{X}$ and $\mathbf{Y}$, respectively. Thus, ($E_{2}$, $\mathbf{y}^{\ast
}$) or just $\mathbf{y}^{\ast }$ asserts ``$E_{2}$ is
performed and $\mathbf{y}^{\ast }$ observed,''  so we may
abbreviate $\mathsf{Infr}_{E_{2}}[(E_{2},\mathbf{y}^{\ast })]$ as $\mathsf{%
Infr}_{E_{2}}[\mathbf{y}^{\ast }]$. Likewise for $\mathbf{x}^{\ast }$. A
generic $\mathbf{z}$ is used when needed.

\subsection{Sufficiency Principle (Weak Likelihood Principle)}\label{sec2.3}

For informative inference about $\theta$ in $E$, if $T_{E}$ is a (minimal)
sufficient statistic for $E$, the Sufficiency Principle asserts the following:
\begin{quote}
SP: If $T_{E}(\mathbf{z}_{1})=T_{E}(
\mathbf{z}_{2})$, then %
$\mathsf{Infr}_{E}[
\mathbf{z}_{1}]=\mathsf{Infr}_{E}[\mathbf{z}_{2}]$.%
\end{quote}
That is, since inference within the model is to be computed using the value
of $T_{E}(\mathbf{\cdot})$ and its sampling distribution, identical values of $%
T_{E}$ have identical inference implications, within the stipulated model.
Nothing in our argument will turn on the minimality requirement, although it
is common.

\subsubsection{\texorpdfstring{Model checking.}{Model checking}}\label{sec2.3.1}

An essential part of the statements of the principles SP, WCP and SLP is
that the validity of the model is granted as adequately representing the
experimental conditions at hand [\citet{Bir62}, page 280]. Thus, accounts that
adhere to the SLP are not thereby prevented from analyzing features of the
data, such as residuals, in checking the validity of the statistical model
itself. There is some ambiguity on this point in Casella and  Berger
(\citeyear{GarJolJon02}):
\begin{quote}
Most model checking is, necessarily, based on statistics other than a
sufficient statistic. For example, it is common practice to examine
residuals from a model\ldots  Such~a practice immediately violates the
Sufficiency Principle, since the \textit{residuals} are not based on
sufficient statistics. (Of course such a practice directly violates the
[strong] LP also.) [Casella and  Berger (\citeyear{GarJolJon02}), pages 295--296.]
\end{quote}

They warn that before considering the SLP and WCP, ``we must
be comfortable with the model'' [\textit{ibid}, page 296]. It seems to us more
accurate to regard the principles as inapplicable, rather than violated,
when the adequacy of the relevant model is lacking. Applying a principle
will always be relative to the associated experimental model.

\subsubsection{Can two become one?}\label{sec2.3.2}

The SP is sometimes called the weak likelihood principle, limited as it is
to a single experiment $E$, with its sampling distribution. This suggests that
if an arbitrary SLP pair, $(E_{1},\mathbf{x}^{\ast})$ and $(E_{2},\mathbf{y}%
^{\ast})$, could be viewed as resulting from a single experiment (e.g., by a
mixture), then perhaps they could become inferentially equivalent using SP.
This will be part of Birnbaum's argument, but is neatly embedded in his
larger gambit to which we now turn.

\subsection{Birnbaumization: Key Gambit in Birnbaum's Argument}\label{sec2.4}

The larger gambit of Birnbaum's argument may be dubbed \textit{Birnbaumization}. An
experiment has been run, label it as $E_{2}$, and $\mathbf{y}^{\ast}$
observed. Suppose, for the parametric inference at hand, that $\mathbf{y}%
^{\ast}$ has an SLP pair $\mathbf{x}^{\ast}$ in a distinct experiment $E_{1}$%
. Birnbaum's task is to show the two are evidentially equivalent, as the SLP
requires.

We are to imagine that performing $E_{2}$ was the result of flipping a fair
coin (or some other randomizer given as irrelevant to $\theta$) to decide
whether to run $E_{1}$ or $E_{2}$. Cox terms this the ``enlarged experiment'' [\citet{Cox78}, page 54], $E_{B}$. We are
then to define a statistic $T_{B}$ that stipulates that if $(E_{2},\mathbf{y}%
^{\ast})$ is observed, its SLP pair $\mathbf{x}^{\ast}$ in the unperformed
experiment is reported;
\begin{eqnarray*}
T_{B}(E_{i},\mathbf{Z}_{i})= \left\{ \begin{array}
{l@{\quad}l} \bigl(E_{1},\mathbf{x}^{\ast}\bigr), & \mbox{if }
\bigl(E_{1},\mathbf{x}^{\ast}\bigr)\mbox{ or }\bigl(E_{2},%
\mathbf{y}^{\ast}\bigr),
\\
(E_{i},\mathbf{z}_{i}), & \mbox{otherwise.}%
\end{array} \right.
\end{eqnarray*}

Birnbaum's argument focuses on the first case and ours will as well.

Following our simplifying notation, whenever $E_{2}$ is performed and $\mathbf{Y}%
=\mathbf{y}^{\ast }$ observed, and $\mathbf{y}^{\ast }$ is seen to admit an
SLP pair, then label its particular SLP pair $(E_{1},\mathbf{x}^{\ast })$.
Any problems of nonuniqueness in identifying SLP pairs are put to one side,
and Birnbaum does not consider them. Thus, when $(E_{2},\mathbf{y}^{\ast })$ is
observed, $T_{B}$ reports it as $(E_{1},\mathbf{x}^{\ast })$. This yields
the Birnbaum experiment, $E_{B}$, with its statistic $T_{B}$. We abbreviate
the inference (about $\theta $) in $E_{B}$ as
\begin{eqnarray*}
\mathsf{Infr}_{E_{B}}\bigl[\mathbf{y}^{\ast }\bigr].
\end{eqnarray*}%
The inference implication (about $\theta $) in $E_{B}$ from $\mathbf{y}%
^{\ast }$ under Birnbaumization is
\begin{eqnarray*}
\bigl(E_{2},\mathbf{y}^{\ast }\bigr)\Rightarrow
\mathsf{Infr}_{E_{B}}\bigl[\mathbf{x}^{\ast
}\bigr],
\end{eqnarray*}%
where the computation in $E_{B}$ is always a convex combination over $E_{1}$
and $E_{2}$. But also,
\begin{eqnarray*}
\bigl(E_{1},\mathbf{x}^{\ast }\bigr)\Rightarrow
\mathsf{Infr}_{E_{B}}\bigl[\mathbf{x}^{\ast
}\bigr].
\end{eqnarray*}%
It follows that, within $E_{B}$, $\mathbf{x}^{\ast }$ and $\mathbf{y}^{\ast
} $ are inferentially equivalent. Call this claim
\begin{eqnarray*}
[B]: \mathsf{Infr}_{E_{B}}[\mathbf{x}^{\ast}]=\mathsf{Infr}_{E_{B}}[\mathbf{%
y}^{\ast}].
\end{eqnarray*}

The argument is to hold for any SLP pair. Now [B] does not yet reach the SLP
which requires
\begin{eqnarray*}
\mathsf{Infr}_{E_{1}}\bigl[\mathbf{x}^{\ast}\bigr]=
\mathsf{Infr}_{E_{2}}\bigl[\mathbf{y}%
^{\ast}\bigr].
\end{eqnarray*}
But Birnbaum does not stop there. Having constructed the hypothetical
experiment $E_{B}$, we are to use the WCP to condition back down to the
known experiment $E_{2}$. But this will not produce the SLP as we now show.

\subsection{Why Appeal to Hypothetical Mixtures?}\label{sec2.5}

Before turning to that, we address a possible query: why suppose the
argument makes any appeal to a hypothetical mixture? (See also Section~\ref{sec5.1}.)
The reason is this: The SLP does not refer to mixtures. It is a universal
generalization claiming to hold for an arbitrary SLP pair. But we have no
objection to imagining [as Birnbaum does (\citeyear{Bir62}), page 284] a universe of all of the
possible SLP pairs, where each pair has resulted from a $\theta $-irrelevant
randomizer (for the given context). Then, when $\mathbf{y}^{\ast }$ is
observed, we pluck the relevant pair and construct $T_{B}$. Our question
is this: why should the inference implication from $\mathbf{y}^{\ast }$ be
obtained by reference to $\mathsf{Infr}_{E_{B}}[\mathbf{y}^{\ast }]$, the
convex combination? Birnbaum does not stop at [B], but appeals to the WCP.
Note the WCP is based on the outcome $\mathbf{y}^{\ast }$ being given.

\section{SLP Violation Pairs}\label{sec3}

Birnbaum's argument is of central interest when we have SLP violations. We
may characterize an SLP violation as any inferential context where the
antecedent of the SLP is true and the consequent is false:
\begin{quote}
SLP violation: $(E_{1},\mathbf{x}^{\ast})$ and $(E_{2},\mathbf{y}%
^{\ast})$ form an SLP pair, but $\mathsf{Infr}_{E_{1}}[\mathbf{x}^{\ast}]\neq%
\mathsf{Infr}_{E_{2}}[\mathbf{y}^{\ast}]$.
\end{quote}

An SLP pair that violates the SLP will be called an \textit{SLP violation pair} (from
$E_{1}$, $E_{2}$, resp.).

It is not always emphasized that whether (and how) an inference method
violates the SLP depends on the type of inference to be made, even within an
account that allows SLP violations. One cannot just look at the data, but
must also consider the inference. For example, there may be no SLP violation
if the focus is on point against point hypotheses, whereas in computing a
statistical significance probability under a null hypothesis there may be.
``Significance testing of a hypothesis\ldots is viewed by
many as a crucial element of statistics, yet it provides a startling and
practically serious example of conflict with the [SLP].''
[Berger and Wolpert (\citeyear{BerWol84}), pages 104--105.] The following is a dramatic example that
often arises in this context.

\subsection{Fixed versus Sequential Sampling}\label{sec3.1}

Suppose $\mathbf{X}$ and $\mathbf{Y}$ are samples from distinct experiments $%
E_{1}$ and $E_{2}$, both distributed as \textsf{N}$(\theta ,\sigma ^{2})$,
with $\sigma ^{2}$ identical and known, and $p$-values are to be calculated
for the null hypothesis $H_{0}$: $\theta =0$ against $H_{1}$: $\theta \neq 0$%
.

In $E_{2}$ the sampling rule is to continue sampling until $\overline{y}%
_{n}>c_{\alpha}=1.96\sigma/\sqrt{n}$, where $\overline{y}_{n}%
=\frac{1}{n}\sum_{i=1}^{n}y_{i}$. In $E_{1}$, the sample size $n$
is fixed and $\alpha=0.05$.

In order to arrive at the SLP pair, we have to consider the particular
outcome observed. Suppose that $E_{2}$ is run and is first able to stop with
$n=169$ trials.  Denote this result as $\mathbf{y}^{\ast }$. A choice for its SLP pair $\mathbf{x}^{\ast }$
would be $(E_{1},1.96\sigma / \sqrt{169})$, and the SLP violation is the fact
that the $p$-values associated with $\mathbf{x}^{\ast }$ and $\mathbf{y}^{\ast }$ differ.
%

\subsection{Frequentist Evidence in the Case of Significance Tests}\label{sec3.2}

\begin{quote}
``[S]topping `when the data looks good' can be a
serious error when combined with frequentist measures of evidence. For
instance, if one used the stopping rule [above]\ldots but analyzed the data
as if a \textit{fixed} sample had been taken, one could \textit{guarantee}
arbitrarily strong frequentist\vadjust{\goodbreak} `significance' against $H_{0}$~\ldots\,.'' [Berger and Wolpert (\citeyear{BerWol84}), page~77.]
\end{quote}

From their perspective, the problem is with the use of frequentist
significance. For a detailed discussion in favor of the irrelevance of this
stopping rule, see Berger and Wolpert (\citeyear{BerWol84}), pages 74--88. For sampling theorists,
by contrast, this example ``taken in the context of
examining consistency with $\theta $ $=0$, is enough to refute the strong
likelihood principle'' [\citet{Cox78}, page 54], since, with
probability $1$, it will stop with a `nominally' significant result even
though $\theta =0$. It contradicts what Cox and Hinkley call
``the weak repeated sampling principle''
[\citet{CoxHin74}, page 51]. More generally, the frequentist principle of
evidence (FEV) would regard small $p$-values as misleading if they result from
a procedure that readily generates small $p$-values under $H_{0}$.\footnote{%
\citeauthor{MayCox06} (\citeyear{MayCox06}), page 254:

FEV: $\mathbf{y}$ is (strong) evidence against $H_{0}$, if and only
if, were $H_{0}$ a correct description of the mechanism generating $\mathbf{y%
}$, then, with high probability this would have resulted in a less
discordant result than is exemplified by $\mathbf{y}$.}

For the sampling theorist, to report a $1.96$ standard deviation difference
known to have come from optional stopping, just the same as if the sample
size had been fixed, is to discard relevant information for inferring
inconsistency with the null, while ``according to any
approach that is in accord with the strong likelihood principle, the fact
that this particular stopping rule has been used is
irrelevant.'' [\citet{CoxHin74}, page 51.]\footnote{%
Analogous situations occur without optional stopping, as with selecting a
data-dependent, maximally likely, alternative [\citet{CoxHin74}, Example~2.4.1, page 51].
See also Mayo and Kruse (\citeyear{ArnSjo01}).} The actual $p$-value will depend
of course on when it stops. We emphasize that our argument does not turn on
accepting a frequentist principle of evidence (FEV), but these
considerations are useful both to motivate and understand the core principle
of Birnbaum's argument, the WCP.

\section{The Weak Conditionality Principle (WCP)}\label{sec4}

From Section~\ref{sec2.4} we have [B] $\mathsf{Infr}_{E_{B}}[\mathbf{x}^{\ast }]\hspace*{-0.75pt}=\hspace*{-0.75pt}%
\mathsf{Infr}_{E_{B}}[\mathbf{y}^{\ast}]$ since the inference implication is
by the constructed $T_{B}$. How might Birnbaum move from [B] to the SLP, for
an arbitrary pair $\mathbf{x}^{\ast}$and $\mathbf{y}^{\ast}$?

There are two possibilities. One would be to insist informative inference
ignore or be insensitive to sampling distributions. But since we know that
SLP violations result because of the difference in sampling distributions,
to simply deny them would obviously render his argument circular (or else
irrelevant for sampling theory). We assume Birnbaum does not intend his
argument to be circular and Birnbaum relies on further steps to which we
now turn.

\subsection{Mixture $(E_{\mathrm{mix}})$: Two Instruments of Different Precisions
\texorpdfstring{[\citet{Cox58}]}{[Cox (1958)]}}\label{sec4.1}

The crucial principle of inference on which Birnbaum's argument rests is the
weak conditionality principle (WCP), intended to indicate the relevant
sampling distribution in the case of certain mixture experiments. The famous
example to which we already alluded, ``is now usually called
the `weighing machine example,' which draws attention to the need for
conditioning, at least in certain types of problems'' [\citet{Rei92}, page 582].

We flip a fair coin to decide which of two instruments, $E_{1}$ or $E_{2}$,
to use in observing a Normally distributed random sample $\mathbf{Z}$ to
make inferences about mean $\theta $. $E_{1}$ has variance of $1$, while
that of $E_{2}$ is $10^{6}$. We limit ourselves to mixtures of two
experiments.

In testing a null hypothesis such as $\theta =0$, the same $\mathbf{z}$
measurement would correspond to a much smaller $p$-value were it to have come
from $E_{1}$ rather than from $E_{2}$: denote them as $p_{1}(\mathbf{z})$
and $p_{2}(\mathbf{z})$, respectively. The overall (or unconditional)
significance level of the mixture $E_{\mathrm{mix}}$ is the convex combination of the
$p$-values: $[p_{1}(\mathbf{z})+p_{2}(\mathbf{z})]/2$. This would give a
misleading report of how precise or stringent the actual experimental
measurement is [Cox and \citet{autokey26}, page 296]. [See Example~4.6, \citet{CoxHin74}, pages 95--96; \citet{Bir62}, page 280.]

Suppose that we know we have observed a measurement from $E_{2}$ with its
much larger variance:
\begin{quote}
The unconditional test says that we can assign this a higher level of
significance than we ordinarily do, because if we were to repeat the
experiment, we might sample some quite different distribution. But this fact
seems irrelevant to the interpretation of an observation which we know came
from a distribution [with the larger variance]. [\citet{Cox58}, page 361.]
\end{quote}

The WCP says simply: \textit{once it is known which} $E_{i}$ \textit{has
produced} $\mathbf{z}$, \textit{the $p$-value or other inferential assessment
should be made with reference to the experiment actually run}.

\subsection{Weak Conditionality Principle (WCP) in the Weighing Machine
Example}\label{sec4.2}

We first state the WCP in relation to this example.

We are given $(E_{\mathrm{mix}},\mathbf{z}_{i})$, that is, $(E_{i},\mathbf{z}_{i})$
results from mixture experiment $E_{\mathrm{mix}}$. WCP exhorts us to condition to be
relevant to the experiment actually producing the outcome. This is an
example of what Cox terms ``conditioning for
relevance.''
\begin{quote}
WCP: Given $(E_{\mathrm{mix}},\mathbf{z}_{i})$, condition on the $E_{i}$
producing the result
\begin{eqnarray*}
&&(E_{\mathrm{mix}},\mathbf{z}_{i})\Rightarrow \mathsf{Infr}_{E_{i}}
\bigl[(E_{\mathrm{mix}},\mathbf{z}%
_{i})\bigr]\\
&&\quad =p_{i}(
\mathbf{z})=\mathsf{Infr}_{E_{i}}[\mathbf{z}_{i}].  \end{eqnarray*}

Do not use the unconditional formulation
\begin{eqnarray*}
&&(E_{\mathrm{mix}},\mathbf{z}_{i})\nRightarrow
\mathsf{Infr}_{E_{\mathrm{mix}}}\bigl[(E_{\mathrm{mix}},%
\mathbf{z}_{i})\bigr]\\
&&\quad =\bigl[p_{1}(\mathbf{z})+p_{2}(
\mathbf{z})\bigr]/2.
\end{eqnarray*}

The concern is that
\begin{eqnarray*}
 \mathsf{Infr}_{E_{\mathrm{mix}}}\bigl[(E_{\mathrm{mix}},
\mathbf{z}_{i})\bigr]=\bigl[p_{1}(\mathbf{z})+p_{2}(%
\mathbf{z})\bigr]/2\neq p_{i}(\mathbf{z}).
\end{eqnarray*}
\end{quote}

There are three sampling distributions, and the WCP says the relevant one to
use whenever\break  $\mathsf{Infr}_{E_{\mathrm{mix}}}[\mathbf{z}_{i}]\neq \mathsf{Infr}%
_{E_{i}}[\mathbf{z}_{i}]$ is the one known to have generated the result
[\citet{Bir62}, page 280]. In other cases the WCP would make no difference.

\subsection{The WCP and Its Corollaries}\label{sec4.3}

We can give a general statement of the WCP as follows:

A mixture $E_{\mathrm{mix}}$ selects between $E_{1}$ and $E_{2}$, using a $\theta $%
-irrelevant process, and it is given that $(E_{i},\mathbf{z}_{i})$ results, $%
i=1,2$. WCP directs the inference implication. Knowing we are
mapping an outcome from a mixture, there is no need to repeat the first
component of $(E_{\mathrm{mix}},\mathbf{z}_{i})$, so it is dropped except when a
reminder seems useful:
\begin{enumerate}[(ii)]
\item[(i)] Condition to obtain relevance:
\begin{eqnarray*}
(E_{\mathrm{mix}},\mathbf{z}_{i})\Rightarrow \mathsf{Infr}_{E_{i}}
\bigl[(E_{\mathrm{mix}},\mathbf{z}%
_{i})\bigr]=
\mathsf{Infr}_{E_{i}}(\mathbf{z}_{i}).
\end{eqnarray*}%
In words, $\mathbf{z}_{i}$ arose from $E_{\mathrm{mix}}$ but the inference
implication is based on $E_{i}$.

\item[(ii)] Eschew unconditional formulations:
\begin{eqnarray*}
(E_{\mathrm{mix}},\mathbf{z}_{i})\nRightarrow\mathsf{Infr}_{E_{\mathrm{mix}}}[
\mathbf{z}%
_{i}],
 \end{eqnarray*}
whenever the unconditional treatment yields a different inference
implication,
\begin{eqnarray*}
\mbox{that is, whenever }\mathsf{Infr}_{E_{\mathrm{mix}}}[
\mathbf{z}_{i}]\neq\mathsf{Infr}%
_{E_{i}}[
\mathbf{z}_{i}].    %
\end{eqnarray*}
\end{enumerate}
\begin{Note*}\hspace*{-8pt} $\mathsf{Infr}_{E_{\mathrm{mix}}}[\mathbf{z}_{i}]$ which abbreviates\break  $%
\mathsf{Infr}_{E_{\mathrm{mix}}}[(E_{\mathrm{mix}},\allowbreak \mathbf{z}_{i})]$ asserts that the
inference implication uses the convex combination of the relevant pair of
experiments.
\end{Note*}

We now highlight some points for reference.

\subsubsection{\texorpdfstring{WCP makes a difference.}{WCP makes a difference}}\label{sec4.3.1}

The cases of interest here are where applying WCP would alter the
unconditional implication. In these cases WCP makes a difference.

Note that (ii) blocks computing the inference implication from $(E_{\mathrm{mix}},%
\mathbf{z}_{i})$ as $\mathsf{Infr}_{E_{\mathrm{mix}}}[\mathbf{z}_{i}]$, whenever $%
\mathsf{Infr}_{E_{\mathrm{mix}}}[\mathbf{z}_{i}]\neq\mathsf{Infr}_{E_{i}}[\mathbf{z}%
_{i}]$ for $i=1,2$. Here $E_{1}$, $E_{2}$ and $E_{\mathrm{mix}}$ would correspond to three
sampling distributions.

WCP requires the experiment and its outcome to be given or known: If it is
given only that $\mathbf{z }$ came from $E_{1}$ or $E_{2}$, and not which,
then WCP does not authorize (i). In fact, we would wish to block such an
inference implication. For instance,
\begin{eqnarray*}
(E_{1}\mbox{ or }E_{2},\mathbf{z})\nRightarrow
\mathsf{Infr}_{E_{1}}[\mathbf{z%
}].
\end{eqnarray*}

Point on notation: The use of ``$\Rightarrow$%
'' is for a given outcome. We may allow it to be used
without ambiguity when only a disjunction is given, because while $E_{1}$
entails ($E_{1}$ or $E_{2}$), the converse does not hold. So no erroneous
substitution into an inference implication would follow.

\subsubsection{\texorpdfstring{Irrelevant augmentation: Keep irrelevant\break facts irrelevant
(Irrel).}{Irrelevant augmentation: Keep irrelevant facts irrelevant
(Irrel)}}\label{sec4.3.2}

Another way to view the WCP is to see it as exhorting us to keep what is
irrelevant to the sampling behavior of the experiment performed irrelevant
(to the inference implication). Consider Birnbaum's (\citeyear{Bir69}), page 119, idea that a
``trivial'' but harmless addition to any given experimental
result $\mathbf{z }$ might be to toss a fair coin and augment $\mathbf{z }$
with a report of heads or tails (where this is irrelevant to the original
model). Note the similarity to attempts to get an exact significance level
in discrete tests, by allowing borderline outcomes to be declared
significant or not (at the given level) according to the outcome of a coin
toss. The WCP, of course, eschews this. But there is a crucial ambiguity to
avoid. It is a harmless addition only if it remains harmless to the
inference implication. If it is allowed to alter the test result, it is
scarcely harmless.

A holder of the WCP may stipulate that a given $\mathbf{z}_{i}$ can always
be augmented with the result of a $\theta$-irrele\-vant randomizer, provided
that it remains irrelevant to the inference implication about $\theta$ in $%
E_{i}$. We can abbreviate this irrelevant augmentation of a given result $%
\mathbf{z}_{i}$ as a conjunction: $(E_{i} \ \&\ \mathsf{Irrel})$,
\begin{quote}
(Irrel): $\mathsf{Infr}_{E_{i}}[(E_{i}\ \&\ \mathsf{Irrel},\mathbf{z%
}_{i})]=\mathsf{Infr}_{E_{i}}[\mathbf{z}_{i}]$, $i=1,2$.
\end{quote}

We illuminate this in the next subsection.

\subsubsection{Is the WCP an equivalence?}\label{sec4.3.3}

``It was the adoption of an unqualified equivalence
formulation of conditionality, and related concepts, which led, in my 1962
paper, to the monster of the likelihood axiom'' [\citet{Kal75N1}, page 263]. He admits the contrast with ``the one-sided form
to which applications'' had been restricted [\citet{Bir69}, page 139,
note 11]. The question of whether the WCP is a proper equivalence relation,
holding in both directions, is one of the most central issues in the
argument. But what would be alleged to be equivalent?

Obviously not the unconditional and the conditional inference implications:
the WCP makes a difference just when they are inequivalent, that is, when $%
\mathsf{Infr}_{E_{\mathrm{mix}}}[\mathbf{z}_{i}]\neq\mathsf{Infr}_{E_{i}}[\mathbf{z}%
_{i}]$. Our answer is that the WCP involves an inequivalence as well as an
equivalence. The WCP prescribes conditioning on the experiment known to have
produced the data, and not the other way around. It is their inequivalence
that gives Cox's WCP its normative proscriptive force. To assume the WCP
identifies $\mathsf{Infr}_{E_{\mathrm{mix}}}[\mathbf{z}_{i}]$ and $\mathsf{Infr}%
_{E_{i}}[\mathbf{z}_{i}]$ leads to trouble. (We return to this in Section \ref{sec7}.)

However, there is an equivalence in WCP (i). Further, once the outcome is given, the
addition of $\theta $-irrelevant features about the selection of the
experiment performed are to remain irrelevant to the inference implication:
\begin{eqnarray*}
\mathsf{Infr}_{E_{i}}\bigl[(E_{\mathrm{mix}},\mathbf{z}_{i})
\bigr]=\mathsf{Infr}%
_{E_{i}}\bigl[(E_{i}\ \&\
\mathsf{Irrel},\mathbf{z}_{i})\bigr].
\end{eqnarray*}
Both are the same as $\mathsf{Infr}_{E_{i}}[\mathbf{z}_{i}]$. While claiming
that $\mathbf{z }$ came from a mixture, even knowing it came from a
nonmixture, may seem unsettling, we grant it for purposes of making out
Birnbaum's argument. By (Irrel), it cannot alter the inference implication
under $E_{i}$.

\section{Birnbaum's SLP Argument}\label{sec5}

\subsection{Birnbaumization and the WCP}\label{sec5.1}

What does the WCP entail as regards Birnbaumization? Now WCP refers to
mixtures, but is the Birnbaum experiment $E_{B}$ a mixture experiment? Not
really. One cannot perform the following: Toss a fair coin (or other $\theta
$-irrelevant randomizer). If it lands heads, perform an experiment $E_{2}$
that yields a member of an SLP pair $\mathbf{y}^{\ast }$; if tails, observe
an experiment that yields the other member of the SLP pair $\mathbf{x}^{\ast
}$. We do not know what outcome would have resulted from the unperformed
experiment, much less that it would be an outcome with a proportional
likelihood to the observed $\mathbf{y}^{\ast }$. There is a single
experiment, and it is stipulated we know which and what its outcome was.
Some have described the Birnbaum experiment as unperformable, or at most a
``mathematical mixture'' rather than an
``experimental mixture'' [\citet{Kal75N2}, pages 252--253]. Birnbaum himself
calls it a ``hypothetical'' mixture [\citet{Bir62}, page 284].

While a holder of the WCP may simply deny its general applicability in
hypothetical experiments, given that Birnbaum's argument has stood for over
fifty years, we wish to give it maximal mileage. Birnbaumization may be
``performed'' in the sense that $T_{B}$ can
be defined for any SLP pair $\mathbf{x}^{\ast}$, $\mathbf{y}^{\ast}$. Refer
back to the hypothetical universe of SLP pairs, each imagined to have been
generated from a $\theta$-irrelevant mixture (Section~\ref{sec2.5}). When we observe $%
\mathbf{y}^{\ast}$ we pluck the $\mathbf{x}^{\ast}$ companion needed for the
argument. In short, we can Birnbaumize an experimental result: Constructing
statistic $T_{B}$ with the derived experiment $E_{B}$ is the
``performance.'' But what cannot shift in
the argument is the stipulation that $E_{i}$ be given or known (as noted in
Section~\ref{sec4.3.1}), that $i$ be fixed. Nor can the meaning of ``given $%
\mathbf{z}^{\ast}$'' shift through the argument, if it is
to be sound.

Given $\mathbf{z}^{\ast }$, the WCP precludes Birnbaumizing. On the other
hand, if the reported $\mathbf{z}^{\ast }$ was the value of $T_{B}$, then we
are given only the disjunction, precluding the computation relevant for $i$
fixed (Section~\ref{sec4.3.1}). Let us consider the components of Birnbaum's argument.

\subsection{Birnbaum's Argument}\label{sec5.2}

$(E_{2},\mathbf{y}^{\ast })$ is given (and it has an SLP pair $\mathbf{x}%
^{\ast }$). The question is to its inferential import. Birnbaum will seek to
show that
\begin{eqnarray*}
\mathsf{Infr}_{E_{2}}\bigl[\mathbf{y}^{\ast }\bigr]=
\mathsf{Infr}_{E_{1}}\bigl[\mathbf{x}%
^{\ast }\bigr].
\end{eqnarray*}%
The value of $T_{B}$ is $(E_{1},\mathbf{x}^{\ast })$. Birnbaumization maps
outcomes into hypothetical mixtures $E_{B}$:
\begin{enumerate}[(2)]
\item[(1)] If the inference implication is by the stipulations of $E_{B}$,
\begin{eqnarray*}
\bigl(E_{2},\mathbf{y}^{\ast }\bigr)\Rightarrow
\mathsf{Infr}_{E_{B}}\bigl[\mathbf{x}^{\ast
}\bigr]=
\mathsf{Infr}_{E_{B}}\bigl[\mathbf{y}^{\ast }\bigr].
\end{eqnarray*}%
Likewise for $(E_{1},\mathbf{x}^{\ast })$. $T_{B}$ is a sufficient statistic
for $E_{B}$ (the conditional distribution of $\mathbf{Z}$ given $T_{B}$ is
independent of $\theta $).

\item[(2)] If the inference implication is by WCP,
\begin{eqnarray*}
\bigl(E_{2},\mathbf{y}^{\ast}\bigr)\nRightarrow
\mathsf{Infr}_{E_{B}}\bigl[\mathbf{y}^{\ast
}\bigr],
\end{eqnarray*}
rather
\begin{eqnarray*}
\bigl(E_{2},\mathbf{y}^{\ast}\bigr)\Rightarrow
\mathsf{Infr}_{E_{2}}\bigl[\mathbf{y}^{\ast }\bigr]%
\end{eqnarray*}
and
\begin{eqnarray*}
 \bigl(E_{1},\mathbf{x}^{\ast}\bigr)\Rightarrow
\mathsf{Infr}_{E_{1}}\bigl[%
\mathbf{x}^{\ast}\bigr].
\end{eqnarray*}
\end{enumerate}

Following the inference implication according to $E_{B}$ in (1) is at odds
with what the WCP stipulates in (2). Given $\mathbf{y}^{\ast }$,
Birnbaumization directs using the convex combination over the components of $%
T_{B}$; WCP eschews doing so. We will not get
\begin{eqnarray*}
\mathsf{Infr}_{E_{1}}\bigl[\mathbf{x}^{\ast }\bigr]=
\mathsf{Infr}_{E_{2}}\bigl[\mathbf{y}%
^{\ast }\bigr].
\end{eqnarray*}

The SLP only seems to follow by the erroneous identity:
\begin{eqnarray*}
\mathsf{Infr}_{E_{B}}\bigl[\mathbf{z}_{i}^{\ast }\bigr]=
\mathsf{Infr}_{E_{i}}\bigl[\mathbf{z%
}_{i}^{\ast }
\bigr]\quad \mbox{for } i=1,2.
\end{eqnarray*}

\subsection{Refuting the Supposition that [(SP and WCP) entails SLP]}\label{sec5.3}

We can uphold both (1) and (2), while at the same time holding the following:
\begin{quote}
(3) $\mathsf{Infr}_{E_{1}}[\mathbf{x}^{\ast}]\neq\mathsf{Infr}_{E_{2}}[%
\mathbf{y}^{\ast}]$.
\end{quote}

Specifically, any case where $\mathbf{x}^{\ast }$ and $\mathbf{y}^{\ast }$
is an SLP violation pair is a case where (3) is true. Since whenever (3)
holds we have a counterexample to the SLP generalization, this demonstrates
that SP and WCP and not-SLP are logically consistent. Thus, so are WCP and
not-SLP. This refutes the supposition that [(SP and WCP) entails SLP] and
also any purported derivation of SLP from WCP alone.\footnote{%
By allowing applications of Birnbaumization and appropriate choices of the
irrelevant randomization probabilities, SP can be weakened to
``mathematical equivalence,'' or even (with
compounded mixtures) omitted so that WCP would entail SLP. See Birnbaum (\citeyear{Bir72}) and Evans,
Fraser and Monette (\citeyear{EvaFraMon86N3}).}

SP is not blocked in (1). The SP is always relative to a model, here $E_{B}$%
. We have the following:
\begin{quote}
$\mathbf{x}^{\ast}$ and $\mathbf{y}^{\ast}
$ are SLP pairs in $E_{B}$, and\\  $
\mathsf{Infr}_{E_{B}}
\bigl[\mathbf{x}^{\ast}\bigr]=\mathsf{Infr}_{E_{B}}\bigl[
\mathbf{y}%
^{\ast}\bigr]$ (i.e., [B] holds).%
\end{quote}

One may allow different contexts to dictate whether or not to condition
[i.e., whether to apply (1) or (2)], but we know of no inference account
that permits, let alone requires, self-contradictions. By noncontradiction,
for any $(E,\mathbf{z})$, $\mathsf{Infr}_{E}[\mathbf{z}]=\mathsf{Infr}_{E}[%
\mathbf{z}]$.
(``$\Rightarrow $'' is a
function from outcomes to inference implications, and $\mathbf{z}=%
\mathbf{z}$, for any~$\mathbf{z}$.)

\textit{Upholding and applying}. This recalls our points in Section~\ref{sec2.1}.
Applying a rule means following its inference directive. We may uphold the
if-then stipulations in (1) and (2), but to apply their competing
implications in a single case is self-contradictory.

\textit{Arguing from a self-contradiction is unsound}. The slogan that
anything follows from a self-contradiction G and not-G is true, since for
any claim C, the following is a logical truth: If G then (if not-G then C).
Two applications of \textit{modus ponens} yield C. One can also derive
not-C! But since G and its denial cannot be simultaneously true, any such
argument is unsound. (A~sound argument must have true premises and be
logically valid.) We know Birnbaum was not intending to argue from a
self-contradiction, but this may inadvertently occur.\looseness=1

\subsection{What if the SLP Pair Arose from an Actual Mixture?}\label{sec5.4}

What if the SLP pair $\mathbf{x}^{\ast }$, $\mathbf{y}^{\ast }$ arose from a
genuine, and not a Birnbaumized, mixture. (Consider fixed versus sequential
sampling, Section~\ref{sec3.1}. Suppose $E_{1}$ fixes $n$ at 169, the coin flip says
perform $E_{2}$, and it happens to stop at $n=169$.) We may allow
that an unconditional formulation may be defined so that
\begin{eqnarray*}
\mathsf{Infr}_{E_{\mathrm{mix}}}\bigl[\mathbf{x}^{\ast }\bigr]=
\mathsf{Infr}_{E_{\mathrm{mix}}}\bigl[\mathbf{y%
}^{\ast }\bigr].
\end{eqnarray*}%
But WCP eschews the unconditional formulation; it says condition on the
experiment known to have produced $\mathbf{z}_{i}$:
\begin{eqnarray*}
\bigl(E_{\mathrm{mix}},\mathbf{z}_{i}^{\ast }\bigr)\Rightarrow
\mathsf{Infr}_{E_{i}}\bigl[\mathbf{z}%
_{i}^{\ast }
\bigr],\quad  i=1,2.
\end{eqnarray*}%
Any SLP violation pair $\mathbf{x}^{\ast },\mathbf{y}^{\ast }$ remains one:\break  $%
\mathsf{Infr}_{E_{1}}[\mathbf{x}^{\ast }]\hspace*{-1pt}\neq \mathsf{Infr}_{E_{2}}[\mathbf{y%
}^{\ast }]$.

\section{Discussion}\label{sec6}

We think a fresh look at this venerable argument is warranted. Wearing a
logician's spectacles and entering the debate outside of the thorny issues
from decades ago may be an advantage.

It must be remembered that the onus is not on someone who questions if the
SLP follows from SP and WCP to provide suitable principles of evidence,
however desirable it might be to have them. The onus is on Birnbaum to show
that for any given $\mathbf{y}^{\ast }$, a member of an SLP pair with $%
\mathbf{x}^{\ast }$, with different probability models $f_{1}(\cdot)$, $f_{2}(\cdot)$%
, that he will be able to derive from SP and WCP, that $\mathbf{x}^{\ast }$
and $\mathbf{y}^{\ast }$ would have the identical inference implications
concerning shared parameter $\theta $. We have shown that SLP violations do
not entail renouncing either the SP or the WCP.

It is no rescue of Birnbaum's argument that a sampling theorist wants
principles in addition to the WCP to direct the relevant sampling
distribution for\vadjust{\goodbreak} inference; indeed, Cox has given others. It was to make the
application of the WCP in his argument as plausible as possible to sampling
theorists that Birnbaum begins with the type of mixture in Cox's (\citeyear{Cox58}) famous
example of instruments $E_{1}$, $E_{2}$ with different precisions.

We do not assume sampling theory, but employ a formulation that avoids
ruling it out in advance. The failure of Birnbaum's argument to reach the
SLP relies only on a correct understanding of the WCP. We may grant that for
any $\mathbf{y}^{\ast }$ its SLP pair could occur in repetitions (and
may\vadjust{\goodbreak}
even be out there as in Section~\ref{sec2.5}). However, the key point of the WCP is
to deny that this fact should alter the inference implication from the known
$\mathbf{y}^{\ast }$. To insist it should is to deny the WCP. Granted, WCP
sought to identify the relevant sampling distribution for inference from a
specified type of mixture, and a known $\mathbf{y}^{\ast }$, but it is
Birnbaum who purports to give an argument that is relevant for a sampling
theorist and for ``approaches which are independent of this
[Bayes'] principle'' [\citet{Bir62}, page 283]. Its implications for
sampling theory is why it was dubbed ``a~landmark in
statistics'' [\citet{autokey34}, page 307].

Let us look at the two statements about inference implications from a given $%
(E_{2},\mathbf{y}^{\ast})$, applying (1) and (2) in Section \ref{sec5.2}:
\begin{eqnarray*}
\bigl(E_{2},\mathbf{y}^{\ast}\bigr)&\Rightarrow&
\mathsf{Infr}_{E_{B}}\bigl[\mathbf{x}^{\ast}\bigr],
\\
\bigl(E_{2},\mathbf{y}^{\ast}\bigr)&\Rightarrow&
\mathsf{Infr}_{E_{2}}\bigl[\mathbf{y}^{\ast}\bigr].
\end{eqnarray*}
Can both be applied in exactly the same model with the same given $\mathbf{z}
$? The answer is yes, so long as the WCP happens to make no difference:
\begin{eqnarray*}
\mathsf{Infr}_{E_{B}}\bigl[\mathbf{z}_{i}^{\ast}\bigr]=
\mathsf{Infr}_{E_{i}}\bigl[\mathbf{z}%
_{i}^{\ast}
\bigr], \quad i=1,2.
\end{eqnarray*}

Now the SLP must be applicable to an arbitrary SLP pair. However, to assume that (1)
and (2) can be consistently applied for any $\mathbf{x}^{\ast },\mathbf{y}%
^{\ast }$ pair would be to assume no SLP violations are possible, which
really would render Birnbaum's argument circular. So from Section \ref{sec5.3}, the choices
are to regard Birnbaum's argument as unsound (arguing from a contradiction)
or circular (assuming what it purports to prove). Neither is satisfactory.
We are left with competing inference implications and no way to get to the
SLP. There is evidence Birnbaum saw the gap in his argument (\cite{Bir72}),
and in the end he held the SLP only restricted to (predesignated) point
against point hypotheses.\footnote{%
This alone would not oust all sampling distributions. Birnbaum's argument,
even were it able to get a foothold, would have to apply further rounds of
conditioning to arrive at the data alone.}

It is not SP and WCP that conflict; the conflict comes from WCP together
with Birnbaumization---understood as both invoking the hypothetical mixture
and erasing the information as to which experiment the data came. If one
Birnbaumizes, one cannot at the same time uphold the ``keep
irrelevants irrelevant'' (Irrel) stipulation of the WCP. So
for any given $(E,\mathbf{z})$ one must choose, and the answer is
straightforward for a holder of the WCP. To paraphrase Cox's (\citeyear{Cox58}), page 361,
objection to unconditional tests:
\begin{quote}
Birnbaumization says that we can assign $\mathbf{y}^{\ast }$ a different
level of significance than we ordinarily do, because one may identify an SLP
pair $\mathbf{x}^{\ast }$ and construct statistic $T_{B}$. But this fact
seems irrelevant to the interpretation of an observation which we know came
from $E_{2}$. To conceal the index, and use the convex combination, would
give a distorted assessment of statistical significance.
\end{quote}

\section{Relation to other Criticisms of Birnbaum}\label{sec7}

A number of critical discussions of the Birnbaum argument and the SLP exist.
While space makes it impossible to discuss them here, we believe the current
analysis cuts through this extremely complex literature. Take, for example,
the most well-known criticisms by \citet{Dur70} and Kalbfleish (\citeyear{Kal75N2}),
discussed in the excellent paper by Evans, Fraser and Monette (\citeyear{EvaFraMon86N3}).
Allowing that any $\mathbf{y}^{\ast}$ may be viewed as having arisen from
Birnbaum's mathematical mixture, they consider the proper order of
application of the principles. If we condition on the given experiment
first, Kalbfleish's revised sufficiency principle is inapplicable, so
Birnbaum's argument fails. On the other hand, Durbin argues, if we reduce to
the minimal sufficient statistic first, then his revised principle of
conditionality cannot be applied. Again Birnbaum's argument fails. So either
way it fails.

Unfortunately, the idea that one must revise the initial principles in order
to block SLP allows downplaying or dismissing these objections as tantamount
to denying SLP at any cost (please see the references\footnote{%
In addition to the authors cited in the manuscript, see especially comments
by Savage,  Cornfield,  Bross,  Pratt,  Dempster et al.  (\citeyear{Savetal62}) on
Birnbaum. For later discussions, see  \citet{Bar75},  \citet{EvaFraMon86N1},
 Berger and Wolpert (\citeyear{BerWol84}), Birnbaum (\citeyear{Bir70N1}, \citeyear{Bir70N2}), \citet{EvaFraMon86N2},
\citet{autokey35} and references therein.}). We can achieve what they wish to
show, without altering principles, and from WCP alone. Given $\mathbf{y}%
^{\ast}$, WCP blocks Birnbaumization; given $\mathbf{y}^{\ast}$ has been
Birnbaumized, the WCP precludes conditioning.

We agree with Evans, Fraser and Monette (\citeyear{EvaFraMon86N3}), page 193, ``that
Birnbaum's use of [the principles] \ldots  are contrary to the intentions of the
principles, as judged by the relevant supporting and motivating examples.
From this viewpoint we can state that the intentions of S and C do not imply
L.'' [Where S, C and L are our SP, WCP and SLP.] Like
Durbin and Kalbfleisch, they offer a choice of modifications of the
principles to block the SLP. These are highly insightful and interesting; we
agree that they highlight a need to be clear on the experimental model at
hand. Still, it is preferable to state the WCP so as to reflect these
``intentions,''  without which it is robbed
of its function. The problem stems from mistaking WCP as the equivalence $%
\mathsf{Infr}_{E_{\mathrm{mix}}}[\mathbf{z}]=\mathsf{Infr}_{E_{i}}[\mathbf{z}]$
(whether the mixture is hypothetical or actual). This is at odds with the
WCP. The puzzle is solved by adequately stating the WCP. Aside from that, we
need only keep the meaning of terms consistent through the argument.

We emphasize that we are neither rejecting the SP nor claiming that it
breaks down, even in the special case $E_{B}$. The sufficiency of $T_{B}$
within $E_{B}$, as a mathematical concept, holds: the value of $T_{B}$
``suffices'' for $\mathsf{Infr}_{E_{B}}[%
\mathbf{y}^{\ast }]$, the inference from the associated convex combination.
Whether reference to hypothetical mixture $E_{B}$ is relevant for inference
from given $\mathbf{y}^{\ast }$ is a distinct question. For an alternative
criticism see \citet{Eva}.

\section{Concluding Remarks}\label{sec8}

An essential component of informative inference for sampling theorists is
the relevant sampling distribution: it is not a separate assessment of
performance, but part of the necessary ingredients of informative inference.
It is this feature that enables sampling theory to have SLP violations
(e.g., in significance testing contexts). Any such SLP violation, according
to Birnbaum's argument, prevents adhering to both SP and WCP. We have shown
that SLP violations do not preclude WCP and SP.

The SLP does not refer to mixtures. But supposing that $(E_{2},\mathbf{y}%
^{\ast })$ is given, Birnbaum asks us to consider that $\mathbf{y}^{\ast }$
could also have resulted from a $\theta $-irrelevant mixture that selects
between $E_{1}$, $E_{2}$. The WCP says this piece of information should be
irrelevant for computing the inference from $(E_{2},\mathbf{y}^{\ast })$
once given. That is, $\mathsf{Infr}_{E_{i}}[(E_{\mathrm{mix}},\mathbf{y}^{\ast })]=%
\mathsf{Infr}_{E_{i}}[\mathbf{y}^{\ast }],\ i=1,2$. It follows that if $\mathsf{Infr}%
_{E_{1}}[\mathbf{x}^{\ast }]\neq \mathsf{Infr}_{E_{2}}[\mathbf{y}^{\ast }]$,
the two remain unequal after the recognition that $\mathbf{y}^{\ast }$ could
have come from the mixture. What was an SLP violation remains one.

Given $\mathbf{y}^{\ast}$, the WCP says do not Birnbaumize. One is free to
do so, but not to simultaneously claim to hold the WCP in relation to the
given $\mathbf{y}^{\ast}$, on pain of logical contradiction. If one does
choose to Birnbaumize, and to construct $T_{B}$, admittedly the known
outcome $\mathbf{y}^{\ast}$ yields the same value of $T_{B}$ as would $%
\mathbf{x}^{\ast}$. Using the sample space of $E_{B}$ yields [B]: $\mathsf{%
Infr}_{E_{B}}[\mathbf{x}^{\ast}]=\mathsf{Infr}_{E_{B}}[\mathbf{y}^{\ast}]$.
This is based on the convex combination of the two experiments and differs
from both $\mathsf{Infr}_{E_{1}}[\mathbf{x}^{\ast}]$ and $\mathsf{Infr}%
_{E_{2}}[\mathbf{y}^{\ast}]$. So again, any SLP violation remains. Granted,
if only the value of $T_{B}$ is given, using $\mathsf{Infr}_{E_{B}}$ may be
appropriate. For then we are given only the disjunction: either $(E_{1},%
\mathbf{x}^{\ast})$ or $(E_{2},\mathbf{y}^{\ast})$. In that case, one is
barred from using the implication from either individual $E_{i}$. A holder
of WCP might put it this way: once $(E,\mathbf{z})$ is given, whether $E$
arose from a $\theta$-irrelevant mixture or was fixed all along should not
matter to the inference, but whether a result was Birnbaumized or not
should, and does, matter.

There is no logical contradiction in holding that if data are analyzed one
way (using the convex combination in $E_{B}$), a given answer results, and
if analyzed another way (via WCP), one gets quite a different result. One may
consistently apply both the $E_{B}$ and the WCP directives to the same
result, in the same experimental model, only in cases where WCP makes no
difference. To claim for any $\mathbf{x}^{\ast }$, $\mathbf{y}^{\ast }$, the
WCP never makes a difference, however, would assume that there can be no SLP
violations, which would make the argument circular.\footnote{%
His argument would then follow the pattern: If there are SLP violations, then
there are no SLP violations. Note that (V implies not-V) is not a logical
contradiction. It is logically equivalent to not-V. Then, Birnbaum's
argument is equivalent to not-V: denying that $\mathbf{x}^{\ast }$, $\mathbf{%
y}^{\ast }$ can give rise to an SLP violation. That would render it circular.%
} Another possibility would be to hold, as Birnbaum ultimately did, that
the SLP is ``clearly plausible'' [\citet{Bir68}, page 301] only in ``the severely restricted case of a
parameter space of just two points'' where these are
predesignated [\citet{Bir69}, page 128]. But that is to relinquish the general
result.\looseness=-1

\section*{Acknowledgments}
I am extremely grateful to David Cox and Aris Spanos for numerous
discussions, corrections and joint work over many years on this and related
foundational issues. I appreciate the careful queries and detailed suggested
improvements on earlier drafts from anonymous referees, from Jean Miller
and Larry Wasserman. My understanding of Birnbaum was greatly facilitated by
philosopher of science, Ronald Giere, who worked with Birnbaum. I'm also
grateful for his gift of some of Birnbaum's original materials and notes.




\begin{thebibliography}{36}

\bibitem[\protect\citeauthoryear{Barndorff-Nielsen}{1975}]{Bar75}
\begin{barticle}[auto:STB|2014/05/26|13:19:10]
\bauthor{\bsnm{Barndorff-Nielsen},~\bfnm{O.}\binits{O.}}
(\byear{1975}).
\btitle{Comments on paper by J. D. Kalbfleisch}.
\bjournal{Biometrika}
\bvolume{62}
\bpages{261--262}.
\end{barticle}
\bptok{imsref}%
\endbibitem

\bibitem[\protect\citeauthoryear{Berger}{1986}]{EvaFraMon86N1}
\begin{barticle}[auto:STB|2014/05/26|13:19:10]
\bauthor{\bsnm{Berger},~\bfnm{James O.}\binits{J. O.}}
(\byear{1986}).
\btitle{Discussion on a paper by
Evans et al. [On principles and arguments to likelihood]}.
\bjournal{Canad. J. Statist.}
\bvolume{14}
\bpages{195--196}.
\end{barticle}
\bptok{imsref}%
\endbibitem

\bibitem[\protect\citeauthoryear{Berger}{2006}]{Ber06}
\begin{barticle}[mr]
\bauthor{\bsnm{Berger},~\bfnm{James O.}\binits{J. O.}}
(\byear{2006}).
\btitle{The case for objective {B}ayesian analysis}.
\bjournal{Bayesian Anal.}
\bvolume{1}
\bpages{385--402}.
\bid{issn={1936-0975}, mr={2221271}}
\end{barticle}
\bptok{imsref}%
\endbibitem

\bibitem[\protect\citeauthoryear{Berger and Wolpert}{1988}]{BerWol84}
\begin{bbook}[mr]
\bauthor{\bsnm{Berger},~\bfnm{James~O.}\binits{J.~O.}} \AND
\bauthor{\bsnm{Wolpert},~\bfnm{Robert~L.}\binits{R.~L.}}
(\byear{1988}).
\btitle{The Likelihood Principle},
\bedition{2nd} ed.
\bseries{Lecture Notes---Monograph Series}
\bvolume{6}.
\bpublisher{IMS},
\blocation{Hayward, CA}.
\bptnote{check year}%
\end{bbook}
\bptok{imsref}%
\endbibitem

\bibitem[\protect\citeauthoryear{Birnbaum}{1962}]{Bir62}
\begin{barticle}[auto:STB|2014/05/26|13:19:10]
\bauthor{\bsnm{Birnbaum},~\bfnm{Allan}\binits{A.}}
(\byear{1962}).
\btitle{On the foundations of statistical inference}.
\bjournal{J. Amer. Statist. Assoc.}
\bvolume{57}
\bpages{269--306}.
\bnote{Reprinted in \textit{Breakthroughs in
Statistics}
\textbf{1}
(S. Kotz and N. Johnson, eds.)
478--518.
Springer,
New York}.
\end{barticle}
\bptok{imsref}%
\endbibitem

\bibitem[\protect\citeauthoryear{Birnbaum}{1968}]{Bir68}
\begin{bincollection}[auto:STB|2014/05/26|13:19:10]
\bauthor{\bsnm{Birnbaum},~\bfnm{A.}\binits{A.}}
(\byear{1968}).
\btitle{Likelihood}.
In \bbooktitle{International Encyclopedia of the Social Sciences}
\bvolume{9}
\bpages{299--301}.
 \bpublisher{Macmillan and the Free Press},
 \blocation{New York}.
\end{bincollection}
\bptok{imsref}%
\endbibitem

\bibitem[\protect\citeauthoryear{Birnbaum}{1969}]{Bir69}
\begin{bincollection}[auto:STB|2014/05/26|13:19:10]
\bauthor{\bsnm{Birnbaum},~\bfnm{A.}\binits{A.}}
(\byear{1969}).
\btitle{Concepts of statistical evidence}.
In \bbooktitle{Philosophy, Science, and Method: Essays in Honor of Ernest Nagel}
(\beditor{\bfnm{S.}\binits{S.}~\bsnm{Morgenbesser}},
\beditor{\bfnm{P.}\binits{P.}~\bsnm{Suppes}} \AND
\beditor{\bfnm{M.~G.}\binits{M.~G.}~\bsnm{White}}, eds.)
\bpages{112--143}.
\bpublisher{St. Martin's Press},
\blocation{New York}.
\end{bincollection}
\bptok{imsref}%
\endbibitem

\bibitem[\protect\citeauthoryear{Birnbaum}{1970a}]{Bir70N1}
\begin{barticle}[auto:STB|2014/05/26|13:19:10]
\bauthor{\bsnm{Birnbaum},~\bfnm{A.}\binits{A.}}
(\byear{1970a}).
\btitle{Statistical methods in scientific inference}.
\bjournal{Nature}
\bvolume{225}
\bpages{1033}.
\end{barticle}
\bptok{imsref}%
\endbibitem

\bibitem[\protect\citeauthoryear{Birnbaum}{1970b}]{Bir70N2}
\begin{barticle}[auto:STB|2014/05/26|13:19:10]
\bauthor{\bsnm{Birnbaum},~\bfnm{A.}\binits{A.}}
(\byear{1970b}).
\btitle{On Durbin's modified principle of conditionality}.
\bjournal{J. Amer. Statist. Assoc.}
\bvolume{65}
\bpages{402--403}.
\end{barticle}
\bptok{imsref}%
\endbibitem

\bibitem[\protect\citeauthoryear{Birnbaum}{1972}]{Bir72}
\begin{barticle}[mr]
\bauthor{\bsnm{Birnbaum},~\bfnm{Allan}\binits{A.}}
(\byear{1972}).
\btitle{More on concepts of statistical evidence}.
\bjournal{J. Amer. Statist. Assoc.}
\bvolume{67}
\bpages{858--861}.
\bid{issn={0162-1459}, mr={0365793}}
\end{barticle}
\bptok{imsref}%
\endbibitem

\bibitem[\protect\citeauthoryear{Birnbaum}{1975}]{Kal75N1}
\begin{barticle}[auto:STB|2014/05/26|13:19:10]
\bauthor{\bsnm{Birnbaum},~\bfnm{A.}\binits{A.}}
(\byear{1975}).
\btitle{Comments on paper by J.~D.~Kalbfleisch}.
\bjournal{Biometrika}
\bvolume{62}
\bpages{262--264}.
\bptnote{check related}%
\end{barticle}
\bptok{imsref}%
\endbibitem

\bibitem[\protect\citeauthoryear{Casella and Berger}{2002}]{GarJolJon02}
\begin{bbook}[auto:STB|2014/05/26|13:19:10]
\bauthor{\bsnm{Casella},~\bfnm{G.}\binits{G.}} \AND
\bauthor{\bsnm{Berger},~\bfnm{R. L.}\binits{R. L.}}
(\byear{2002}).
\btitle{Statistical Inference},
\bedition{2nd} ed.
\bpublisher{Duxbury Press},
\blocation{Belmont, CA}.
\end{bbook}
\bptok{imsref}%
\endbibitem

\bibitem[\protect\citeauthoryear{Cox}{1958}]{Cox58}
\begin{barticle}[mr]
\bauthor{\bsnm{Cox},~\bfnm{D.~R.}\binits{D.~R.}}
(\byear{1958}).
\btitle{Some problems connected with statistical inference}.
\bjournal{Ann. Math. Statist.}
\bvolume{29}
\bpages{357--372}.
\bid{issn={0003-4851}, mr={0094890}}
\end{barticle}
\bptok{imsref}%
\endbibitem

\bibitem[\protect\citeauthoryear{Cox}{1977}]{Cox77}
\begin{barticle}[mr]
\bauthor{\bsnm{Cox},~\bfnm{D.~R.}\binits{D.~R.}}
(\byear{1977}).
\btitle{The role of significance tests}.
\bjournal{Scand. J. Stat.}
\bvolume{4}
\bpages{49--70}.
\bid{issn={0303-6898}, mr={0448666}}
\bptnote{check related}%
\end{barticle}
\bptok{imsref}%
\endbibitem

\bibitem[\protect\citeauthoryear{Cox}{1978}]{Cox78}
\begin{barticle}[mr]
\bauthor{\bsnm{Cox},~\bfnm{D.~R.}\binits{D.~R.}}
(\byear{1978}).
\btitle{Foundations of statistical inference: The case for eclecticism}.
\bjournal{Aust. N. Z. J. Stat.}
\bvolume{20}
\bpages{43--59}.
\bid{issn={0004-9581}, mr={0501453}}
\end{barticle}
\bptok{imsref}%
\endbibitem

\bibitem[\protect\citeauthoryear{Cox and Hinkley}{1974}]{CoxHin74}
\begin{bbook}[mr]
\bauthor{\bsnm{Cox},~\bfnm{D.~R.}\binits{D.~R.}} \AND
\bauthor{\bsnm{Hinkley},~\bfnm{D.~V.}\binits{D.~V.}}
(\byear{1974}).
\btitle{Theoretical Statistics}.
\bpublisher{Chapman \& Hall},
\blocation{London}.
\bid{mr={0370837}}
\end{bbook}
\bptok{imsref}%
\endbibitem

\bibitem[\protect\citeauthoryear{Cox and Mayo}{2010}]{autokey17}
\begin{bincollection}[auto:STB|2014/05/26|13:19:10]
\bauthor{\bsnm{Cox},~\bfnm{D.~R.}\binits{D.~R.}} \AND
\bauthor{\bsnm{Mayo},~\bfnm{D.~G.}\binits{D.~G.}}
(\byear{2010}).
\btitle{Objectivity
and conditionality in frequentist inference}.
In \bbooktitle{Error and Inference: Recent Exchanges on Experimental Reasoning,
Reliability, and the Objectivity and Rationality of Science}
(\beditor{\bfnm{G.}\binits{G.}~\bsnm{Mayo}} \AND
\beditor{\bfnm{Aris}\binits{A.}~\bsnm{Spanos}}, eds.)
\bpages{276--304}.
\bpublisher{Cambridge Univ. Press},
\blocation{Cambridge}.
\end{bincollection}
\bptok{imsref}%
\endbibitem


\bibitem[\protect\citeauthoryear{Dawid}{1986}]{EvaFraMon86N2}
\begin{barticle}[auto:STB|2014/05/26|13:19:10]
\bauthor{\bsnm{Dawid},~\bfnm{A. P.}\binits{A. P.}}
(\byear{1986}).
\btitle{Discussion on a paper by
Evans et al. [On principles and arguments to likelihood]}.
\bjournal{Canad. J. Statist.}
\bvolume{14}
\bpages{196--197}.
\end{barticle}
\bptok{imsref}%
\endbibitem

\bibitem[\protect\citeauthoryear{Durbin}{1970}]{Dur70}
\begin{barticle}[auto:STB|2014/05/26|13:19:10]
\bauthor{\bsnm{Durbin},~\bfnm{J.}\binits{J.}}
(\byear{1970}).
\btitle{On Birnbaum's theorem on the relation between sufficiency, conditionality and likelihood}.
\bjournal{J. Amer. Statist. Assoc.}
\bvolume{65}
\bpages{395--398}.
\end{barticle}
\bptok{imsref}%
\endbibitem

\bibitem[\protect\citeauthoryear{Evans}{2013}]{Eva}
\begin{bmisc}[auto:STB|2014/05/26|13:19:10]
\bauthor{\bsnm{Evans},~\bfnm{M.}\binits{M.}}
(\byear{2013}). \bhowpublished{What does the proof of Birnbaum's theorem prove? Unpublished manuscript}.
\end{bmisc}
\bptok{imsref}%
\endbibitem



\bibitem[\protect\citeauthoryear{Evans, Fraser and Monette}{1986}]{EvaFraMon86N3}
\begin{barticle}[mr]
\bauthor{\bsnm{Evans},~\bfnm{Michael~J.}\binits{M.~J.}},
\bauthor{\bsnm{Fraser},~\bfnm{Donald~A.~S.}\binits{D.~A.~S.}} \AND
\bauthor{\bsnm{Monette},~\bfnm{Georges}\binits{G.}}
(\byear{1986}).
\btitle{On principles and arguments to likelihood}.
\bjournal{Canad. J. Statist.}
\bvolume{14}
\bpages{181--199}.
\bid{doi={10.2307/3314794}, issn={0319-5724}, mr={0859631}}
\bptnote{check related}%
\end{barticle}
\bptok{imsref}%
\endbibitem

\bibitem[\protect\citeauthoryear{Ghosh, Delampady and Samanta}{2006}]{GhoDelSam06}
\begin{bbook}[mr]
\bauthor{\bsnm{Ghosh},~\bfnm{Jayanta~K.}\binits{J.~K.}},
\bauthor{\bsnm{Delampady},~\bfnm{Mohan}\binits{M.}} \AND
\bauthor{\bsnm{Samanta},~\bfnm{Tapas}\binits{T.}}
(\byear{2006}).
\btitle{An Introduction to {B}ayesian Analysis. Theory and Methods}.
\bseries{Springer Texts in Statistics}.
\bpublisher{Springer},
\blocation{New York}.
\bid{mr={2247439}}
\end{bbook}
\bptok{imsref}%
\endbibitem

\bibitem[\protect\citeauthoryear{Kalbfleisch}{1975}]{Kal75N2}
\begin{barticle}[mr]
\bauthor{\bsnm{Kalbfleisch},~\bfnm{John~D.}\binits{J.~D.}}
(\byear{1975}).
\btitle{Sufficiency and conditionality}.
\bjournal{Biometrika}
\bvolume{62}
\bpages{251--268}.
\bid{issn={0006-3444}, mr={0386075}}
\bptnote{check related}%
\end{barticle}
\bptok{imsref}%
\endbibitem

\bibitem[\protect\citeauthoryear{Lehmann and Romano}{2005}]{LehRom05}
\begin{bbook}[mr]
\bauthor{\bsnm{Lehmann},~\bfnm{E.~L.}\binits{E.~L.}} \AND
\bauthor{\bsnm{Romano},~\bfnm{Joseph~P.}\binits{J.~P.}}
(\byear{2005}).
\btitle{Testing Statistical Hypotheses},
\bedition{3rd} ed.
\bseries{Springer Texts in Statistics}.
\bpublisher{Springer},
\blocation{New York}.
\bid{mr={2135927}}
\end{bbook}
\bptok{imsref}%
\endbibitem

\bibitem[\protect\citeauthoryear{Mayo}{1996}]{May96}
\begin{bbook}[auto:STB|2014/05/26|13:19:10]
\bauthor{\bsnm{Mayo},~\bfnm{D.~G.}\binits{D.~G.}}
(\byear{1996}).
\btitle{Error and the Growth of Experimental Knowledge}.
\bpublisher{Univ. Chicago Press},
\blocation{Chicago, IL}.
\end{bbook}
\bptok{imsref}%
\endbibitem


\bibitem[\protect\citeauthoryear{Mayo}{2010}]{autokey26}
\begin{bincollection}[auto:STB|2014/05/28|10:36:42]
\bauthor{\bsnm{Mayo},~\bfnm{D. G.}\binits{D. G.}}
(\byear{2010}).
\btitle{An error in the argument from
conditionality and sufficiency to the likelihood
principle}.
In \bbooktitle{Error and Inference: Recent
Exchanges on Experimental Reasoning, Reliability, and the Objectivity and
Rationality of Science}
(\beditor{\bfnm{Deborah G.}\binits{D.~G.}~\bsnm{Mayo}}
 \AND
\beditor{\bfnm{Aris}\binits{A.}~\bsnm{Spanos}}, eds.)
\bpages{305--314}.
\bpublisher{Cambridge Univ. Press},
\blocation{Cambridge}.
\end{bincollection}
\bptok{imsref}%
\endbibitem


\bibitem[\protect\citeauthoryear{Mayo and Cox}{2010}]{MayCox06}
\begin{bincollection}[auto:STB|2014/05/28|10:36:42]
\bauthor{\bsnm{Mayo},~\bfnm{Deborah~G.}\binits{D.~G.}} \AND
\bauthor{\bsnm{Cox},~\bfnm{D.~R.}\binits{D.~R.}}
(\byear{2010}).
\btitle{Frequentist statistics as a theory of inductive inference}.
In \bbooktitle{Error and Inference: Recent Exchanges on Experimental Reasoning,
Reliability, and the Objectivity and Rationality of Science}
(\beditor{\bfnm{Deborah G.}\binits{D. G.}~\bsnm{Mayo}}
 \AND
\beditor{\bfnm{Aris}\binits{A.}~\bsnm{Spanos}}, eds.)
\bpages{247--274}.
\bpublisher{Cambridge Univ. Press},
\blocation{Cambridge}.
\bnote{First published in \textit{The Second Erich L. Lehmann Symposium: Optimality} \textbf{49}
(2006) (J. Rojo, ed.) 77--97. \textit{Lecture Notes---Monograph Series}. IMS,
Beachwood, OH}.
\end{bincollection}
\bptok{imsref}%
\endbibitem

\bibitem[\protect\citeauthoryear{Mayo and Cox}{2011}]{MayCox11}
\begin{bincollection}[auto:STB|2014/05/26|13:19:10]
\bauthor{\bsnm{Mayo},~\bfnm{D.~G.}\binits{D.~G.}} \AND
\bauthor{\bsnm{Cox},~\bfnm{D.~R.}\binits{D.~R.}}
(\byear{2011}).
\btitle{Statistical scientist meets a philosopher of science: A conversation}.
In \bbooktitle{Rationality, Markets and Morals: Studies at the Intersection of Philosophy and Economics}
\bvolume{2}
(\beditor{\bfnm{Deborah G.}\binits{D. G.}~\bsnm{Mayo}},
\beditor{\bfnm{Aris}\binits{A.}~\bsnm{Spanos}}  \AND
\beditor{\bfnm{Kent W.}\binits{K. W.}~\bsnm{Staley}}, eds.)
\bnote{(\textit{Special Topic: Statistical Science and Philosophy of Science: Where do (should) They Meet in 2011 and Beyond}?)
(October 18)}
\bpages{103--114}.
\bpublisher{Frankfurt School},
\blocation{Frankfurt}.
\end{bincollection}
\bptok{imsref}%
\endbibitem

\bibitem[\protect\citeauthoryear{Mayo and Kruse}{2001}]{ArnSjo01}
\begin{bincollection}[auto:STB|2014/05/26|13:19:10]
\bauthor{\bsnm{Mayo},~\bfnm{D. G.}\binits{D. G.}} \AND
\bauthor{\bsnm{Kruse},~\bfnm{M.}\binits{M.}}
(\byear{2001}).
\btitle{Principles of
inference and their consequences}.
In \bbooktitle{Foundations
of Bayesianism}
(\beditor{\bfnm{David}\binits{D.}~\bsnm{Corfield}}
\AND
\beditor{\bfnm{Jon}\binits{J.}~\bsnm{Williamson}}, eds.)
\bvolume{24}
\bpages{381--403}.
\bnote{\textit{Applied Logic}}.
\bpublisher{Kluwer Academic Publishers},
\blocation{Dordrecht}.
\end{bincollection}
\bptok{imsref}%
\endbibitem

\bibitem[\protect\citeauthoryear{Mayo and Spanos}{2006}]{MaySpa06}
\begin{barticle}[mr]
\bauthor{\bsnm{Mayo},~\bfnm{Deborah~G.}\binits{D.~G.}} \AND
\bauthor{\bsnm{Spanos},~\bfnm{Aris}\binits{A.}}
(\byear{2006}).
\btitle{Severe testing as a basic concept in a {N}eyman--{P}earson philosophy of induction}.
\bjournal{British J. Philos. Sci.}
\bvolume{57}
\bpages{323--357}.
\bid{doi={10.1093/bjps/axl003}, issn={0007-0882}, mr={2249183}}
\end{barticle}
\bptok{imsref}%
\endbibitem



\bibitem[\protect\citeauthoryear{Mayo and Spanos}{2011}]{MaySpa11}
\begin{bincollection}[auto:STB|2014/05/26|13:19:10]
\bauthor{\bsnm{Mayo},~\bfnm{D.~G.}\binits{D.~G.}} \AND
\bauthor{\bsnm{Spanos},~\bfnm{A.}\binits{A.}}
(\byear{2011}).
\btitle{Error statistics}.
In \bbooktitle{Philosophy of Statistics}
\bvolume{7}
(\beditor{\bfnm{Prasanta S.}\binits{P. S.}~\bsnm{Bandyopadhyay}}
\AND
\beditor{\bfnm{Malcom R.}\binits{M. R.}~\bsnm{Forster}}, eds.)
\bpages{152--198}.
\bnote{\textit{Handbook of the Philosophy of Science}}.
\bpublisher{Elsevier},
\blocation{Amsterdam}.
\end{bincollection}
\bptok{imsref}%
\endbibitem\vadjust{\goodbreak}

\bibitem[\protect\citeauthoryear{Reid}{1992}]{Rei92}
\begin{bincollection}[auto:STB|2014/05/26|13:19:10]
\bauthor{\bsnm{Reid},~\bfnm{N.}\binits{N.}}
(\byear{1992}).
\btitle{Introduction to Fraser (1966) structural probability and a generalization}.
In \bbooktitle{\textit{Breakthroughs in Statistics}}
(\beditor{\bfnm{Samuel}\binits{S.}~\bsnm{Kotz}} \AND
\beditor{\bfnm{Norman~L.}\binits{N.~L.}~\bsnm{Johnson}}, eds.)
\bpages{579--586}.
\bnote{\textit{Springer Series in Statistics}}.
\bpublisher{Springer},
\blocation{New York}.
\end{bincollection}
\bptok{imsref}%
\endbibitem

\bibitem[\protect\citeauthoryear{Savage}{1962a}]{autokey33}
\begin{bbook}[auto:STB|2014/05/26|13:19:10]
\beditor{\bsnm{Savage},~\bfnm{L. J.}\binits{L. J.}}, ed.
(\byear{1962a}).
\btitle{The Foundations of Statistical Inference: A Discussion}.
\bpublisher{Methuen},
\blocation{London}.
\end{bbook}
\bptok{imsref}%
\endbibitem

\bibitem[\protect\citeauthoryear{Savage}{1962b}]{autokey34}
\begin{barticle}[auto:STB|2014/05/26|13:19:10]
\bauthor{\bsnm{Savage},~\bfnm{L.~J.}\binits{L.~J.}}
(\byear{1962b}).
\btitle{Discussion on a paper by A. Birnbaum [On the foundations of statistical inference]}.
\bjournal{J. Amer. Statist. Assoc.}
\bvolume{57}
\bpages{307--308}.
\end{barticle}
\bptok{imsref}%
\endbibitem

\bibitem[\protect\citeauthoryear{Savage}{1970}]{autokey35}
\begin{barticle}[auto:STB|2014/05/26|13:19:10]
\bauthor{\bsnm{Savage},~\bfnm{L.~J.}\binits{L.~J.}}
(\byear{1970}).
\btitle{Comments on a weakened principle of conditionality}.
\bjournal{J. Amer. Statist. Assoc.}
\bvolume{65} (329)
\bpages{399--401}.
\end{barticle}
\bptok{imsref}%
\endbibitem

\bibitem[\protect\citeauthoryear{Savage et~al.}{1962}]{Savetal62}
\begin{barticle}[auto:STB|2014/05/26|13:19:10]
\bauthor{\bsnm{Savage},~\bfnm{L.~J.}\binits{L.~J.}},
\bauthor{\bsnm{Barnard},~\bfnm{G.}\binits{G.}},
\bauthor{\bsnm{Cornfield},~\bfnm{J.}\binits{J.}},
\bauthor{\bsnm{Bross},~\bfnm{I.}\binits{I.}},
\bauthor{\bsnm{Box},~\bfnm{G.~E.~P.}\binits{G.~E.~P.}},
\bauthor{\bsnm{Good},~\bfnm{I.~J.}\binits{I.~J.}},
\bauthor{\bsnm{Lindley},~\bfnm{D. V.}\binits{D. V.}}
et al.
(\byear{1962}).
\btitle{On the foundations of statistical inference: Discussion}.
\bjournal{J. Amer. Statist. Assoc.}
\bvolume{57}
\bpages{307--326}.
\end{barticle}
\bptok{imsref}%
\endbibitem

\end{thebibliography}
\end{document}